\documentclass[3p,times,procedia]{elsarticle}
\usepackage{ecrc}

\volume{00}

\firstpage{1}

\journalname{Transportation Research Procedia}

\runauth{J. Zhang, D. Britto, M. Chraibi, R. L\"ohner, E. Haug and B. Gawenat }

\jid{aaspro}

\usepackage{amssymb}

\biboptions{authoryear}

\usepackage[figuresright]{rotating}

\begin{document}

\begin{frontmatter}

\dochead{The Conference in Pedestrian and Evacuation Dynamics 2014 (PED2014)}

\title{Quantitative validation of PEDFLOW for description of unidirectional pedestrian dynamics}

\author[a]{J. Zhang\corref{cor1}}
\author[b]{D. Britto}
\author[a]{M. Chraibi}
\author[b]{R. L\"ohner}
\author[b]{E. Haug}
\author[b]{B. Gawenat}

\address[a]{J\"ulich Supercomputing Centre, Forschungszentrum J\"ulich GmbH, J¨¹lich 52428, Germany}
\address[b]{Institute for Scientific Architecture, SL Rasch, Leinfelden-Echterdingen 70771,Germany}

\begin{abstract}

The results of a systematic quantitative validation of PEDFLOW based on the experimental data from FZJ are presented.  Unidirectional flow experiments, totaling 28 different combinations with varying entry, corridor and exit widths, were considered. The condition imposed on PEDFLOW was that all the cases should be run with the same input parameters. The exit times and fundamental diagrams for the measuring region were evaluated and compared. This validation process led to modifications and enhancements of the model underlying PEDFLOW. The preliminary conclusions indicate that the results agree well for densities smaller than 3 $m^{-2}$ and a good agreement is observed even at high densities for the corridors with $b_{cor} =2.4~m$, and $b_{cor}=3.0~m$. For densities between 1 and 2  $m^{-2}$ the specific flow and velocities are underpredicted by PEDFLOW. \bigskip
\end{abstract}

\cortext[cor1]{Corresponding author. Tel.: +49-246-161-96554 ; fax: +49-246-161-6656.}
\end{frontmatter}

\email{ju.zhang@fz-juelich.de}

\section{Introduction}
\label{main}

During the last few decades, research on pedestrian and traffic flow has attracted a lot of attention (\cite{Schreckenberg2002, Appert-Rolland2009, Bandini2010,Klingsch2010,Schadschneider2009c,Schadschneider2009}).  The investigation of pedestrian motion plays an important role in guaranteeing the safety of pedestrians in complex buildings or at mass events. A large number of models have been developed in the past and most of them are able to reproduce phenomena of pedestrian movement qualitatively. Before using a model to predict quantitative results like the total evacuation time, it needs to be calibrated thoroughly and quantitatively using empirical data. However, this is still difficult due to a lack of reliable experimental data as well as the surprisingly large differences in available datasets (\cite{Seyfried2009}). Even for the fundamental diagram which states the relationship between pedestrian flow and density, obvious discrepancies can be seen in the literature and handbooks. By comparing the density-velocity or density-specific flow relationships from different researchers like \cite{Fruin1971}, \cite{Predtechenskii1978}, \cite{Weidmann1993} and \cite{Helbing2007}, it can be found that the density $\rho_0$, where the velocity approaches zero due to overcrowding, ranges from 3.8 to 10 $m^{-2}$, while the density $\rho_c$ where the pedestrian flux reaches its maximum ranges from 1.75 to 7 $m^{-2}$. Since these data come from different conditions and different measurement methods, it is not possible to conclude the reasons for the differences. For the calibration of models, it is difficult to set the same conditions to make comparisons.
In this paper we validate the software PEDFLOW used by SL-Rasch Special and Lightweight Structures GmbH in Germany on the basis of the empirical data from well-controlled laboratory experiments. The remainder of the paper is organized as follows: In section 2, we describe the model implemented in the software. The validation process and results are shown in section 3. Thereafter, the conclusions from our validation are discussed.

\section{The PEDFLOW model}

The modeling of pedestrian motion has been the focus of research and development for more than two decades. If one is only interested in average quantities (average density, velocity), continuum models (\cite{Hughes2003}) are an options. For problems requiring more realism, approaches that model each individual are required (\cite{Thalmann2007}). Among these, discrete space models (such as cellular automata (\cite{Blue1998, Teknomo2000, Dijkstra2002a, Schadschneider2002, Kessel2002, Kluepfel2003, Langston2006a}, force-based models (such as the social force model (\cite{Helbing1995,Helbing2002, Lakoba2005}) and agent-based techniques (\cite{Pelechano2006, Badler2008,Guy2009, Guy2010, Torrens2011,Curtis2012}) have been explored extensively. Together with insights from psychology and neuroscience (e.g. \cite{Vishton1995, Torrens2011}), it has become clear that any pedestrian motion algorithm that attempts to model reality should be able to mirror the following empirically known facts and behaviors:
\begin{itemize}
\item Newton's laws of motion apply to humans as well: from one instant to another, we can only move within certain bounds of acceleration, velocity and space;
\item Contact between individuals occurs for high densities; these forces have to be taken into account;
\item Humans have a mental map and plan on how they desire to move globally (e.g. first go here, then there, etc.);
\item In even moderately crowded situations ( $<$ 1 $m^{-2}$ ), humans have a visual horizon of 2.5-5.0 $m$, and a perception range of 120 degrees; thus, the influence of other humans beyond these thresholds is minimal;
\item Humans have a `personal comfort zone'; It is dependent on culture and varies from individual to individual, but it cannot be ignored;
\item Humans walk comfortably at roughly 2 paces per second (frequency: 2 $Hz$); they are able to change the frequency for short periods of time, but will return to 2 $Hz$ whenever possible.
\end{itemize}

The PEDFLOW model (\cite{Loehner2010}) incorporates these requirements as follows: individuals move according to Newton's laws of motion; they follow (via will forces) `global movement targets'; At the local movement level, the motion also considers the presence of other individuals or obstacles via avoidance forces (also a type of will force) and, if applicable, contact forces. PEDFLOW also incorporates a number of psychological factors that, among the many tried over the years, have emerged as important for realistic simulations. Among these, we mention:

\begin{itemize}
\item Determination/Pushiness: it is an everyday experience that in crowds, some people exhibit a more polite behavior than others. This is modeled in PEDFLOW by reducing the collision avoidance forces of more determined or `pushier' individuals. Defining a determination or pushiness parameter p, the avoidance forces are reduced by (1-p).
\item Comfort zone: in some cultures (northern Europeans are a good example) pedestrians want to remain at some minimum distance from contacting others. This comfort zone is an input parameter in PEDFLOW, and is added to the radii of the pedestrians when computing collisions avoidance and pre-contact forces.
\end{itemize}

\section{Model calibration and analysis}
A series of well-controlled experiments have been performed. The pedestrian trajectories were extracted from video recordings semi automatically and with high precision using the software $PeTrack$ (\cite{Boltes2010}). All the validation work was based on trajectories from simulations and experiments. Throughout the validation process, the density-velocity and density-specific flow relationships were compared. Due to its high precision, the Voronoi method was used for measuring these values. The same size and location of the measurement area are adopted for the experimental and computational datasets so as to exclude their influence on the results. Moreover, in order to decrease the scattering of data for both experimental and simulation data, a time period was defined in such a way that the effects of the starting and ending conditions of the observed run are minimal and fluctuations are low. This was done by inspecting the time series of density and velocity. Details for the measurement method can be found in \cite{Steffen2010a} and \cite{ZhangJ2011}.

\subsection{Validation scenarios}

\begin{figure}
\centering{
\includegraphics[height =0.3\textwidth]{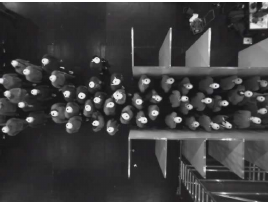}
\includegraphics[height =0.3\textwidth]{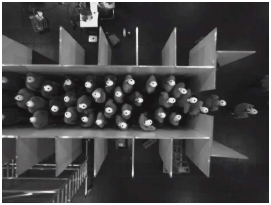}}
\caption{\label{fig-snap} Snapshots from one run of the experiment.}
\end{figure}

In this study, the PEDFLOW simulation software tool was calibrated with data from the experiment of unidirectional flow in straight corridors. Fig. \ref{fig-snap} shows two snapshots from one experimental run. A total of 28 runs in corridors with widths of $b_{cor}=1.8~m$, 2.4 $m$ and 3.0 $m$ were performed. To regulate the pedestrian density in the corridors the widths of the entrance $b_{entrance}$ and the exit $b_{exit}$ were changed in each run. In this way the in- and outflow of the corridors are controlled by the entrance and exit. The details of exit and entrance width in the experiment setting can be found in \cite{ZhangJ2011}. At the beginning of each run, the participants were held within a waiting area. When the experiment starts, they passed through a 4 m passage into the corridor. The passage was used as a buffer to minimize the effect of the entrance. Therefore, the pedestrian flow in the corridor was nearly homogeneous over its entire width. The focus of the study was on the motion dynamics in the 8 m long corridor. An average free velocity $v_0 = 1.55 \pm 0.18~m/s$ was obtained by measuring the free movement of 42 participants.

\subsection{Simulation configuration}

\begin{figure}
\centering{
\includegraphics[height =0.25\textwidth]{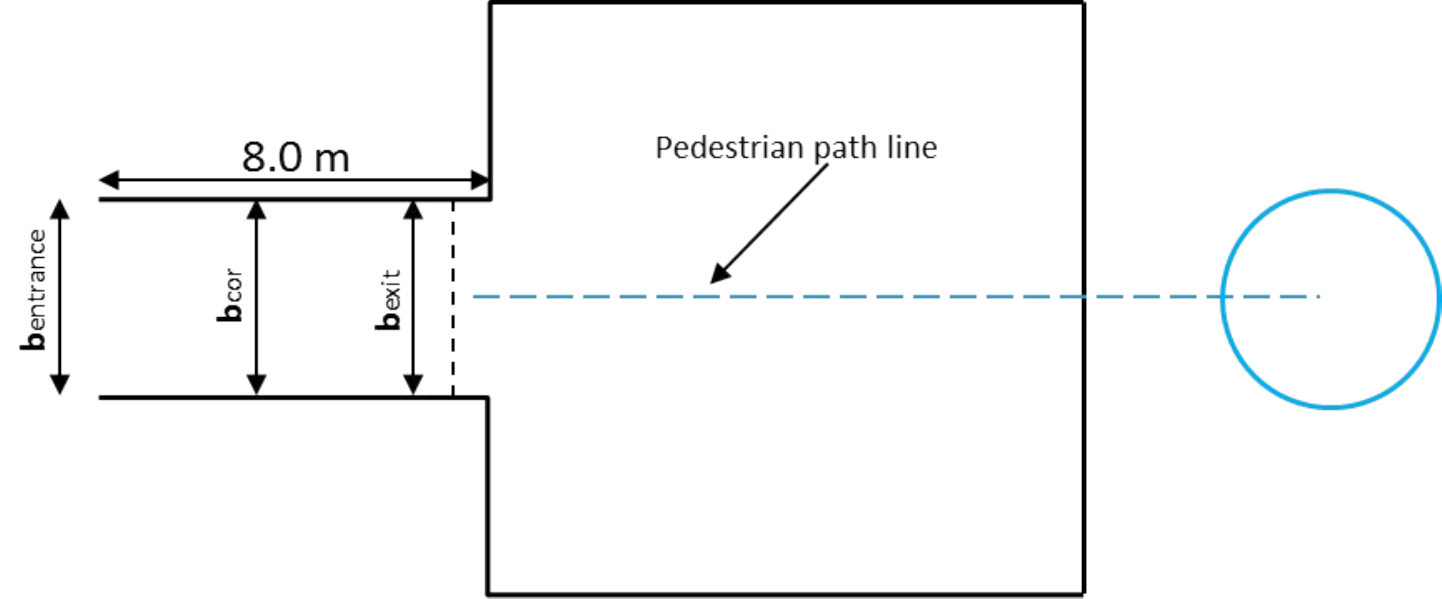}
\includegraphics[height =0.25\textwidth]{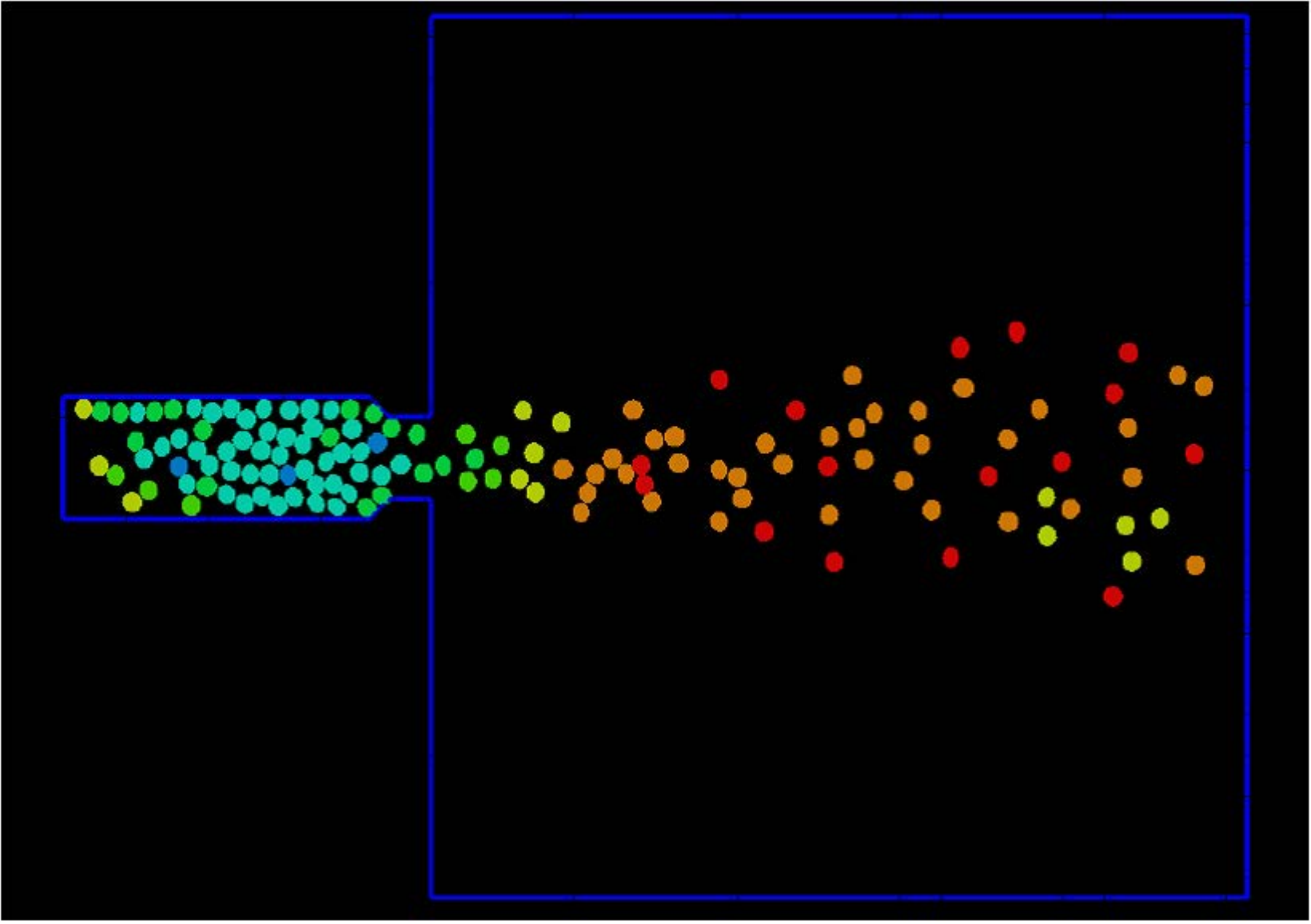}}
\caption{\label{fig-sim} The sketch of the simulation scenario setup and a snapshot of the simulation result.}
\end{figure}

In the simulation, identical geometrical set-ups as in the experiment, including corridor widths and exit widths were implemented. The sketch of the setup and a snapshot from the simulation can be seen in Fig. \ref{fig-sim}.  The difference from the simulation and experimental setup is that the inflow rate into the corridor is not controlled by the entrance width in simulations. At the beginning of the corridor pedestrians are generated based on the timely varying flux data from the experiment. In this way it is possible to dispense with the density-triggering by the waiting area, and a further source of errors is eliminated. The parameters used in the simulation are shown in Table 1.

\begin{table}[h]
\caption{Parameters set up used in the simulation.}
\begin{tabular*}{\hsize}{@{\extracolsep{\fill}}llll@{}}
\toprule
Parameters & Value & Parameters & Value\\
\colrule
itype	& 1	& radi &	0.24\\
mtype	& 0	& vrad	& 0.1\\
otype	& 1	& ellmi/mx	& 0.50/1.00\\
col	& 1	& pshmi/mx	&0.00/0.00\\
velo	& 1.60 $\pm$ 0.01 $m/s$	&czone	&0.15\\
trlx	& 0.50 $\pm$ 0.05& &		\\
\botrule
\end{tabular*}
\end{table}

\subsection{Validation results}
\begin{figure}
\centering{
\includegraphics[height =0.40\textwidth]{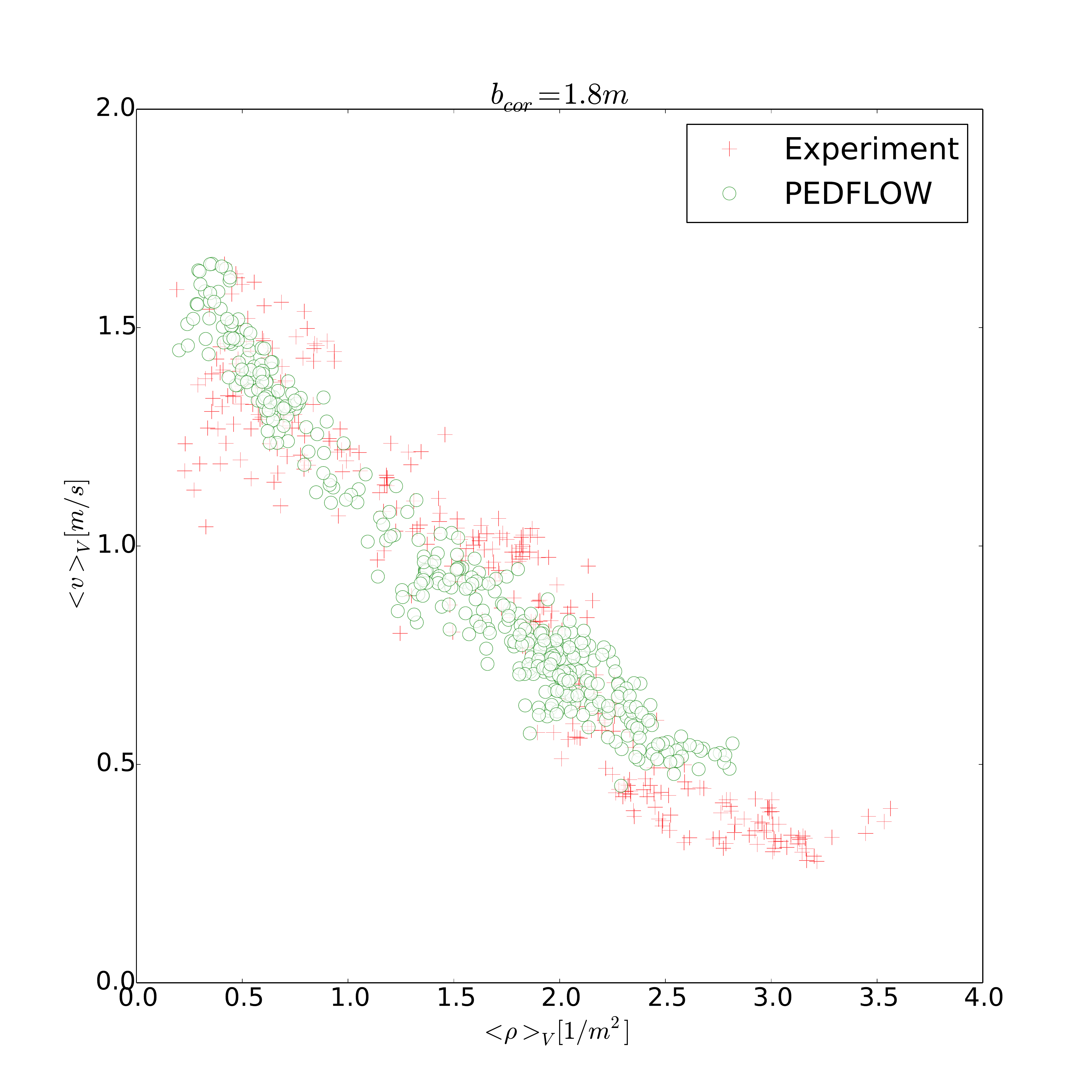}
\includegraphics[height =0.40\textwidth]{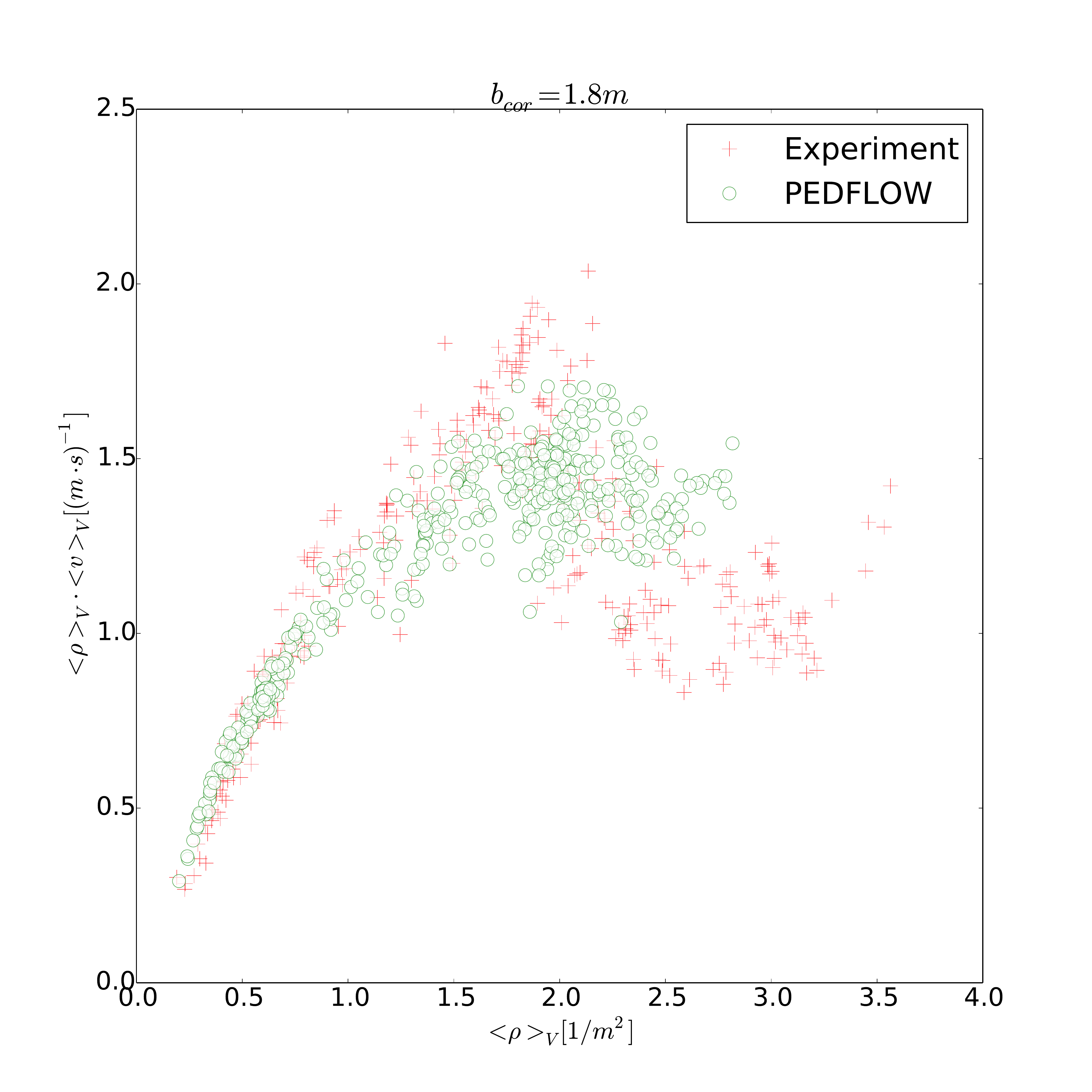}
\includegraphics[height =0.40\textwidth]{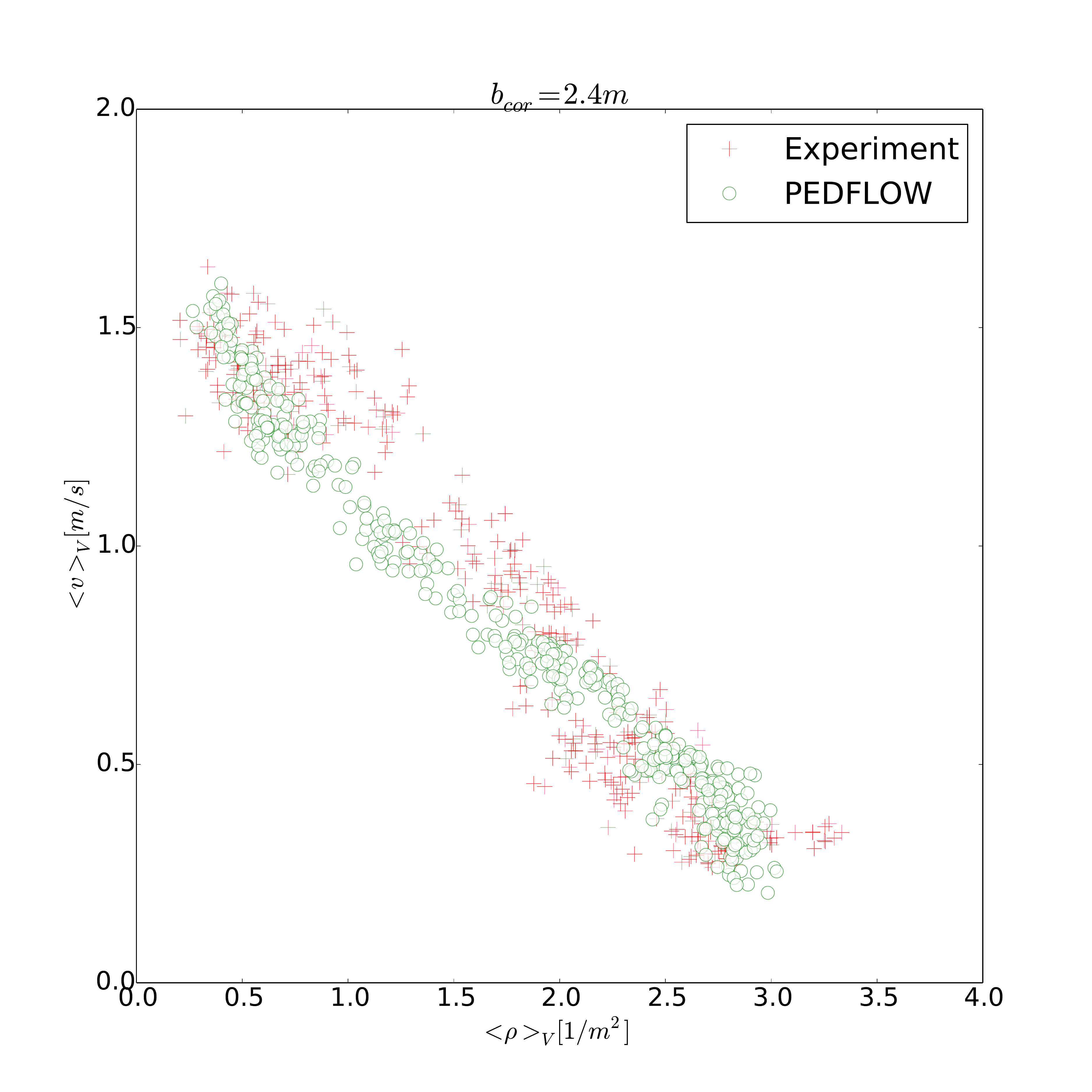}
\includegraphics[height =0.40\textwidth]{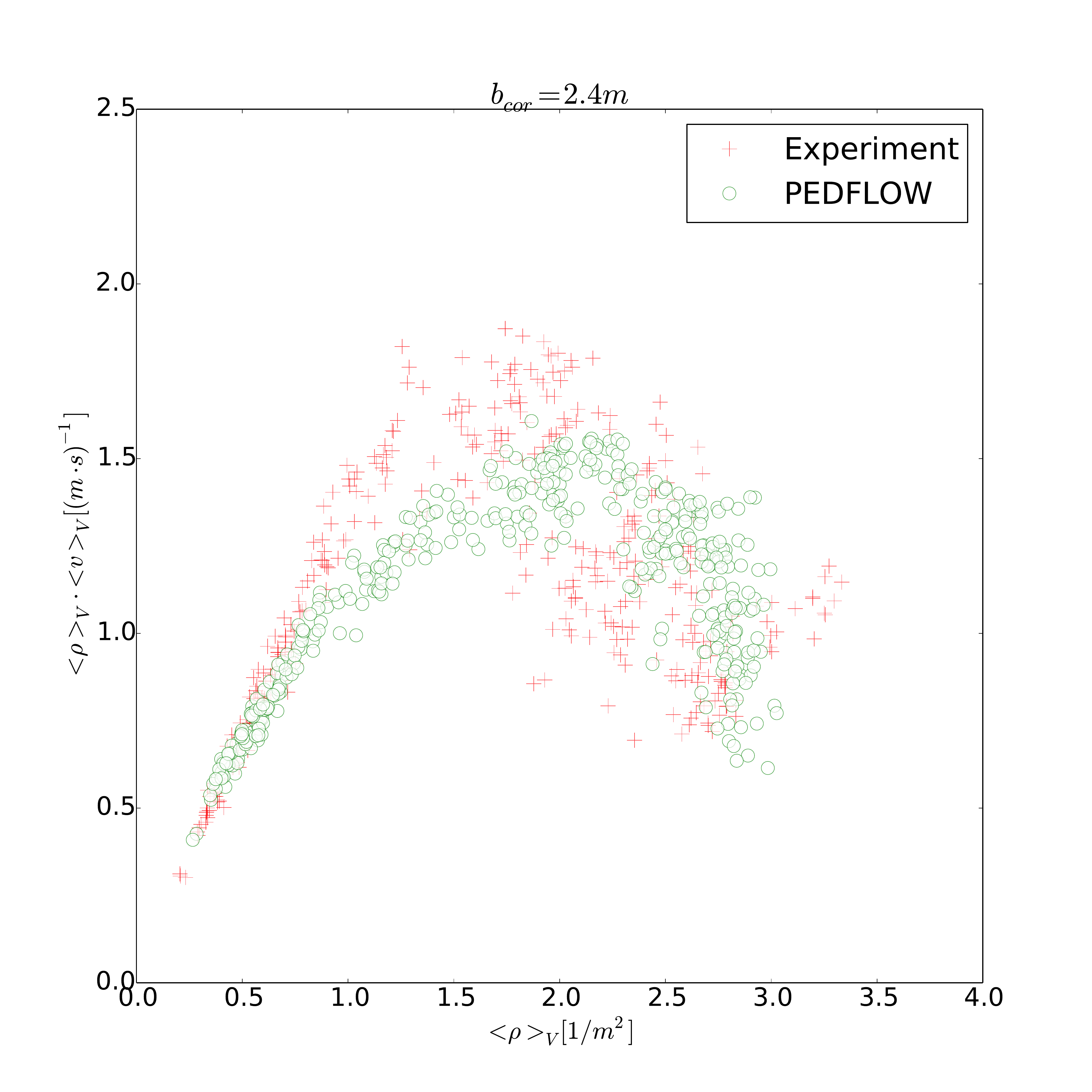}
\includegraphics[height =0.40\textwidth]{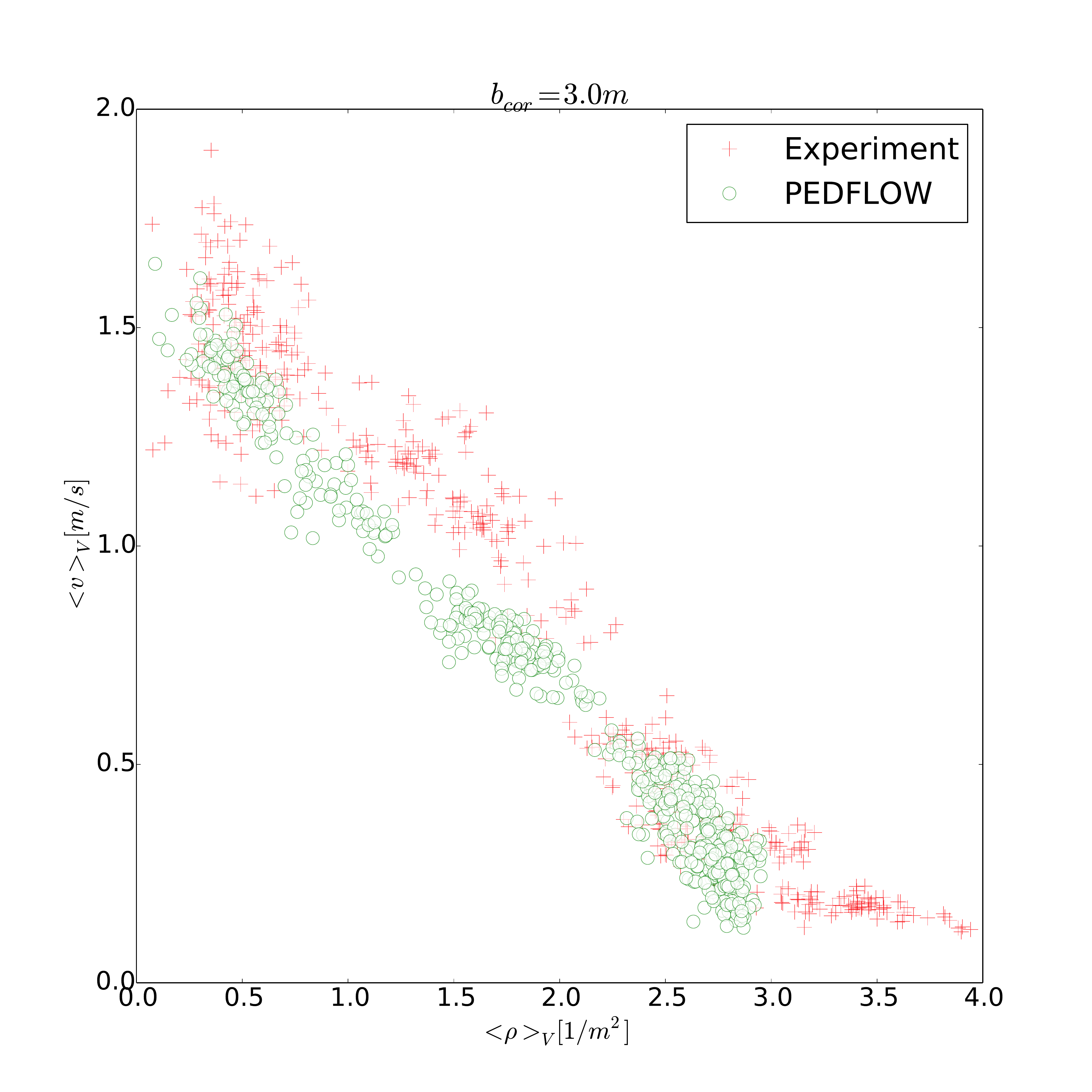}
\includegraphics[height =0.40\textwidth]{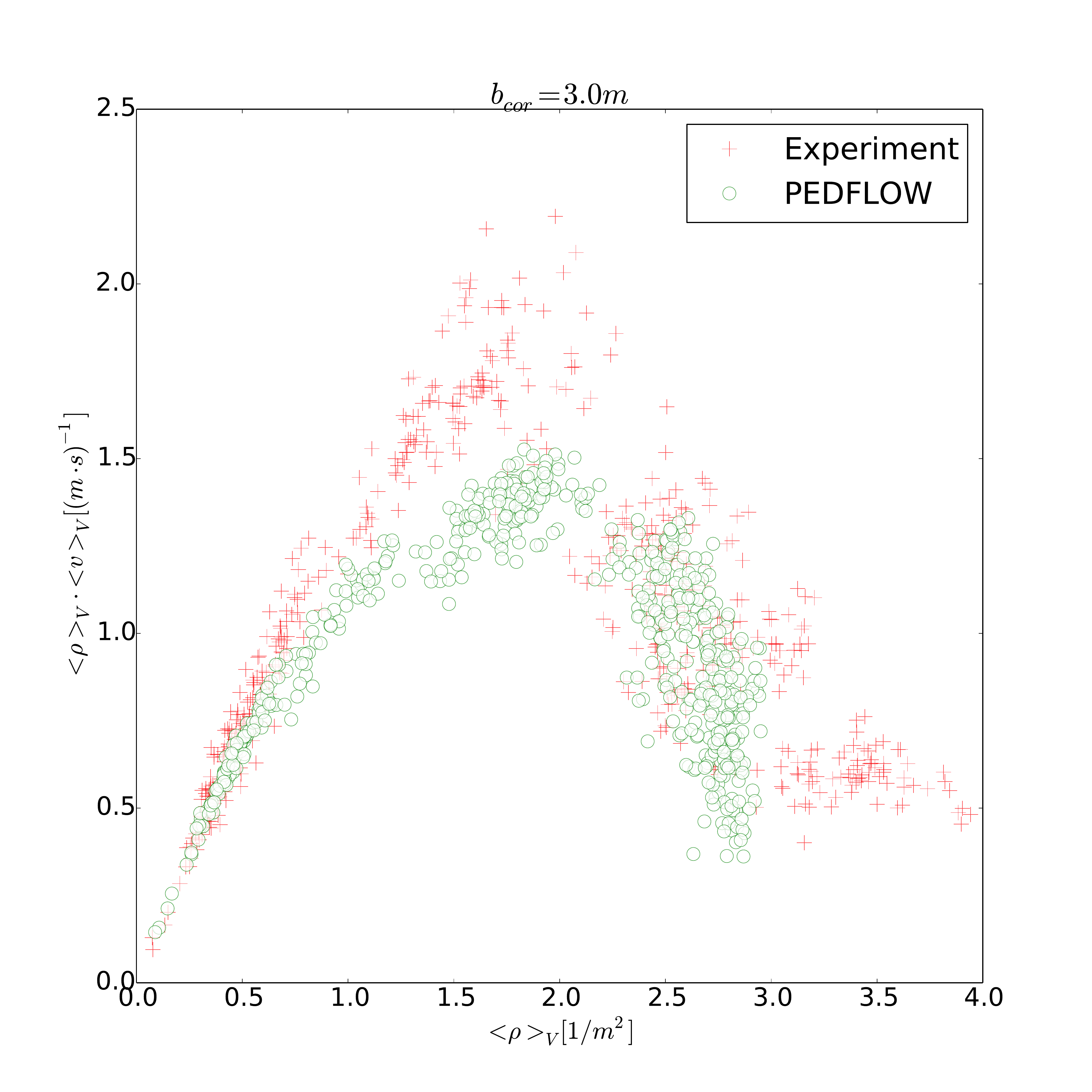}}
\caption{\label{fig-comp} Comparison of experimental and PEDFLOW results for $b_{cor} = 1.8~m, 2.4~m$ and 3.0 $m$.}
\end{figure}

This validation process led to modifications and enhancements of the model underlying PEDFLOW. After several enhancements of the model, the simulation results show a good agreement with the experiment results. Fig. \ref{fig-comp} shows the comparison of the fundamental diagrams from experiments and PEDFLOW for 1.8 $m$, 2.4 $m$ and 3.0 $m$ wide corridors. It found that the results agree well for densities smaller than 3 $m^{-2}$ and a good agreement is observed even at high densities for the corridors with $b_{cor} = 2.4~m$, and $b_{cor} = 3.0~m$. For densities between 1 and 2 $m^{-2}$ the specific flow and velocities are under-predicted by PEDFLOW.
The simulation has shown that a user of PEDFLOW is able to reproduce a decrease of the velocity with the density with one set of parameters. The simulation results agree well with the experimental ones for densities smaller than 3 $m^{-2}$.

\section{Summary}
In this paper, the model underlying the Software PEDFLOW that is used by SL-Rasch Special and Lightweight Structures GmbH in Germany was validated. The validation process was based on the well-controlled laboratory experiments in straight corridors. Unidirectional pedestrian flows in a straight corridor with different widths were selected as the calibration scenarios.  Since one is able to obtain a high precision for the trajectories in both the experiments and the simulations, the validations were mainly made using the trajectories. This validation-process led to modifications and enhancements of the model underlying PEDFLOW and the simulation has shown that a user of PEDFLOW is able to reproduce a decrease of the velocity with the density with one set of parameters. The simulation results agree well with the experimental ones for densities smaller than 3 $m^{-2}$.

\end{document}